\documentclass[a4paper,final]{article}
\pdfoutput=1

\usepackage{xparse}
\usepackage{xpatch}
\usepackage{etoolbox}
\usepackage{environ}
\usepackage{lineno}

\usepackage[hmarginratio=1:1,top=20mm,left=22mm,columnsep=20pt]{geometry} %

\usepackage{lmodern}%
\usepackage[T1]{fontenc}
\usepackage[utf8]{inputenc}

\usepackage{scrextend}
\changefontsizes[10pt]{8pt}

\usepackage[scaled=.98,sups,osf]{XCharter}
\usepackage[scaled=0.86]{berasans}
\usepackage[scaled=1.03]{inconsolata}

\usepackage{setspace}
\RequirePackage[
	final,
	tracking=smallcaps, 
	letterspace=20
]{microtype}

\usepackage{titlesec}
\titleformat{\section}[block]
	{\Large\mdseries\scshape\centering}
	{\thesection.}{.3em}
	{\MakeLowercase}
\titleformat{\subsection}[block]
	{\Large\mdseries\scshape}
	{\thesubsection.}{.3em}
	{\MakeLowercase}
\titleformat{\subsubsection}[block]
	{\large\mdseries\scshape}
	{\thesubsubsection.}{.3em}
	{\MakeLowercase}
\titleformat{\paragraph}[runin]
	{\large\mdseries\scshape}
	{\thesubsubsection.}{.3em}
	{\MakeLowercase}
\titleformat{\subparagraph}[runin]
	{\large\mdseries\scshape}
	{\thesubsubsection.}{.3em}
	{\MakeLowercase}

\usepackage{xcolor}
\usepackage[breakable]{tcolorbox}

\colorlet{keyword}{blue!25!black}
\colorlet{literal}{red!40!black}

\usepackage[british]{babel}
\usepackage{csquotes}
\usepackage[all]{foreign}
\defasforeign[etsimilia]{et similia}
\defasforeign[Etsimilia]{Et similia}
\defnotforeign[wrt]{{\xperiodafter{w.r.t}}}
\defnotforeign[wl]{{\xperiodafter{w.l.o.g}}}

\usepackage{amsmath}
\usepackage{amsfonts,amssymb,bm,stmaryrd}
\usepackage{mathtools}
\usepackage{esvect}

\setdisplayskipstretch{1.2}
\renewcommand{\arraystretch}{1.2}
\SetSymbolFont{stmry}{bold}{U}{stmry}{m}{n}
\allowdisplaybreaks

\usepackage{amsthm,thmtools}

\declaretheoremstyle[%
	postheadspace = .5em,
	headfont=\normalfont\large\mdseries\scshape,
	notefont=\normalfont, notebraces={(}{)},
	bodyfont=\normalfont\itshape,
	headformat={\MakeLowercase{\expandafter\NAME{}} \NUMBER \NOTE},
]{plain}

\declaretheoremstyle[
	postheadspace = .5em,
	headfont=\normalfont\large\mdseries\scshape,
	notefont=\normalfont, notebraces={(}{)},
	bodyfont=\normalfont\itshape,
	headformat={\MakeLowercase{\expandafter\NAME{}} \NUMBER \NOTE},
]{definition}

\declaretheoremstyle[
	postheadspace = .5em,
	headfont=\normalfont\large\mdseries\scshape,
	notefont=\normalfont, notebraces={(}{)},
	bodyfont=\normalfont\itshape,
	headformat={\MakeLowercase{\expandafter\NAME{}} \NOTE},
]{notation}

\declaretheoremstyle[
	postheadspace = .5em,
	headfont=\normalfont\large\mdseries\scshape,
	notefont=\normalfont, notebraces={(}{)},
	bodyfont=\normalfont,
	headformat={\MakeLowercase{\expandafter\NAME{}} \NUMBER \NOTE},
	qed={$\triangleleft$}
]{remark}

\declaretheoremstyle[
	postheadspace = .5em,
	headfont=\normalfont\large\mdseries\scshape,
	notefont=\normalfont, notebraces={(}{)},
	bodyfont=\normalfont,
	headformat={\MakeLowercase{\expandafter\NAME{}}\xspace\NOTE},
	qed={$\Box$}
]{proof}

\makeatletter
\newcommand{\declarecrefname}[5]{
	\Crefname{#1}{#4}{#5}
	\if@cref@capitalise
		\crefname{#1}{#4}{#5}	
	\else
		\crefname{#1}{#2}{#3}	
	\fi
}
\newcommand\usecrefname[1]{\noexpand\noexpand\noexpand\csuse{cref@#1@name}}
\makeatother

\declaretheorem[
  name=\usecrefname{theorem},
  parent=section,
]{theorem}

\declaretheorem[
  name=\usecrefname{corollary},
  sibling=theorem
]{corollary}

\declaretheorem[
  name=\usecrefname{lemma},
  sibling=theorem
]{lemma}

\declaretheorem[
  name=\usecrefname{definition},
  parent=section,
  style=definition
]{definition}

\declaretheorem[
  name=\usecrefname{remark},
  parent=section,
  style=remark,
]{remark}

\declaretheorem[
  name=\usecrefname{example},
  sibling=remark,
]{example}

\declaretheorem[
  name=proof,
  style=proof,
  numbered=no,
]{proof}

\usepackage{float}
\usepackage{caption}

\let\realfigurename=\figurename
\let\realtablename=\tablename
\usepackage[section]{placeins}

\usepackage[inline]{enumitem}
\setlist{itemsep=1pt,topsep=1pt}

\usepackage{tikz}
\usetikzlibrary{calc,arrows,positioning}

\newcommand*\autoflowbox[1]{#1}
\newcommand*\autoflowbreak{}
\providecommand*\afbox\autoflowbox
\makeatletter
\newcommand*\autoflow@sep{\hspace{3ex}}
\newcommand*\autoflow@linestart{\hfill}
\newcommand*\autoflow@lineend{\hfill}
\newcommand*\autoflow@firstlinestart{\autoflow@linestart}
\newcommand*\autoflow@lastlineend{\autoflow@lineend}
\newcommand*\autoflow@boxwrapper[1]{#1}
\newcommand*\autoflow@makeline[1]{\makebox[\linewidth][l]{#1}}

\newlength{\autoflow@befreskip}
\newlength{\autoflow@middleskip}
\newlength{\autoflow@afterskip}
\newlength{\autoflow@tmplinewidth}

\newenvironment{autoflow}
{%
	\begingroup%
	\newif\ifautoflow@breakpending%
	\autoflow@breakpendingfalse%
	\def\autoflow@currentline{\autoflow@firstlinestart}%
	\def\autoflow@sep@{}%
	\def\autoflowbreak{\autoflow@breakpendingtrue}%
	\def\autoflowbox##1{%
		\ifautoflow@breakpending%
			\autoflow@makeline{\autoflow@currentline\autoflow@lineend}%
			\vspace{\autoflow@middleskip}%
			\def\autoflow@currentline{\autoflow@linestart\autoflow@boxwrapper{##1}}%
			\autoflow@breakpendingfalse%
		\else%
			\let\autoflow@tmpline\autoflow@currentline%
			\addto\autoflow@tmpline{\autoflow@sep@\autoflow@boxwrapper{##1}}%
			\def\autoflow@sep@{\autoflow@sep}%
			\settowidth{\autoflow@tmplinewidth}{\autoflow@tmpline}%
			\ifnum\linewidth>\autoflow@tmplinewidth\relax%
				\let\autoflow@currentline\autoflow@tmpline%
			\else%
				\autoflow@makeline{\autoflow@currentline\autoflow@lineend}%
				\vspace{\autoflow@middleskip}%
				\def\autoflow@currentline{\autoflow@linestart\autoflow@boxwrapper{##1}}%
			\fi%
		\fi%
	}%
	\parskip=0pt%
	\parindent=0pt%
	\par%
	\vspace{\autoflow@befreskip}%
	\begin{minipage}{\linewidth}\ignorespaces%
}{%
	\autoflow@makeline{\autoflow@currentline\autoflow@lastlineend}%
	\end{minipage}%
	\endgroup%
	\vspace{\autoflow@afterskip}%
	\noindent\ignorespacesafterend%
}
\makeatother

\usepackage{hyperref}
\usepackage{nameref}

\usepackage[capitalise,nameinlink,noabbrev]{cleveref}

\makeatletter
\newcommand{\customlabel}[4][0]{%
	\protected@write\@auxout{}{\string\newlabel{#3}{{#4}{\thepage}{#4}{#3}{}}}%
	\protected@write\@auxout{}{\string\newlabel{#3@cref}{{[#2][#1][#1]#4}{\thepage}}}%
}
\newcommand{\crefv}[1]{%
	\begingroup\@cref@compressfalse\@cref@sortfalse\cref{#1}\endgroup%
}
\newcommand{\Crefv}[1]{%
	\begingroup\@cref@compressfalse\@cref@sortfalse\Cref{#1}\endgroup%
}
\makeatother

\hypersetup{
	hidelinks,
	final,
}

\usepackage[
	sectionbib,
	square,
	numbers,
	sort&compress
	]{natbib}
\newcommand{\defeq}{\triangleq}

\newenvironment{snippetbox}[1][]{%
	\begin{tcolorbox}[
		colback=black!5,colframe=black!30,
		breakable,
		boxsep=3pt,
		boxrule=.5pt,
		sharp corners,
		#1
	]}{%
		\end{tcolorbox}\noindent\ignorespacesafterend%
	}
\newenvironment{snippet}[1][]{%
	\begin{snippetbox}%
	\begingroup%
	\renewcommand*{\\}[1][]{\endmath\par\math\displaystyle}%
	\begin{internallinenumbers}%
	\resetlinenumber[1]%
	\modulolinenumbers[1]%
	\setlength\linenumbersep{5pt}%
	\math\displaystyle%
}{%
	\endmath%
	\end{internallinenumbers}%
	\endgroup%
	\end{snippetbox}%
	\noindent\ignorespacesafterend%
}
\newenvironment{snippet*}[1][]{%
	\begin{snippetbox}[#1]%
	\begingroup%
	\renewcommand*{\\}[1][]{\endmath\par\math\displaystyle}%
	\math\displaystyle%
}{%
	\endmath%
	\endgroup%
	\end{snippetbox}%
	\noindent\ignorespacesafterend%
}

\newcommand*{\code}[1]{\text{\ttfamily\upshape #1}}
\newcommand*{\keyword}[1]{{\text{\color{keyword}\sffamily\upshape #1}}}
\newcommand*{\literal}[1]{{\text{\color{literal}\ttfamily\upshape #1}}}
\newcommand*{\conc}{\keyword{;\kern1pt}}
\newcommand*{\comarrow}{\mathrel{\keyword{\ttfamily-\kern-.45ex>}}}
\newcommand*{\actorimpl}{\mathrel{\color{keyword}\triangleright}}
\newcommand*{\pid}[1]{\ensuremath{\mathsf{#1}}}
\newcommand*{\aspid}[1]{\IfBooleanF{#1}{\pid}}
\newcommand*{\precongr}{\preceq}
\newcommand*{\congr}{\equiv}
\newcommand*{\nil}{\boldsymbol 0}
\newcommand*{\Eta}{\ensuremath{H}}
\makeatletter
\newcommand*\com{%
	\@ifstar%
	\@com%
	\@@com%
}
\newcommand*\@com[2]{#1 \comarrow #2}
\newcommand*\@@com[4]{\@com{\pid{#1}.{\ensuremath{#2}}}{\pid{#3}.{\ensuremath{#4}}}}

\newcommand{\mcomldelim}{\{}
\newcommand{\mcomrdelim}{\}}
\newcommand{\mcommdelim}{|}
\NewDocumentCommand\mcom{s m o}{{\mathchoice%
	{\IfBooleanTF{#1}{\@mcom{#2}[#3]}{\@@mcom{#2}[#3]}}% displaystyle
	{\@mcom{#2}[#3]}{\@mcom{#2}[#3]}{\@mcom{#2}[#3]} % not display
}}
\NewDocumentCommand\@mcom{m o}{{%
	\renewcommand*\\[1][]{,\,}%
	\renewcommand*\vdots\dots%
	\left\mcomldelim#1%
	\IfNoValueF{#2}{\,\middle\mcommdelim\,#2}%
	\right\mcomrdelim%
}}
\NewDocumentCommand\@@mcom{m o}{\begingroup%
	\renewcommand*\dots\vdots%
	\def\arraystretch{1}%
	\left\mcomldelim\hskip-1ex\begin{array}{c}#1\end{array}%
	\IfNoValueF{#2}{\hskip-1ex\middle\mcommdelim\hskip-1ex\begin{array}{c}#2\end{array}}%
	\hskip-1ex\right\mcomrdelim%
\endgroup}
\makeatother
\let\msel\mcom

\NewDocumentCommand\sel{s m m m}{%
	\com*{\aspid{#1}{#2}}{\aspid{#1}{#3}[#4]}}
\makeatletter
\newcommand*\cond{%
	\@ifstar%
	\@cond%
	\@@cond%
}
\newcommand*\@cond[3]{\keyword{if\;}#1\keyword{\;then\;}#2\keyword{\;else\;}#3}
\newcommand*\@@cond[4]{\@cond{\pid{#1}.{\ensuremath{#2}}}{#3}{#4}}
\makeatother
\newcommand{\rec}[3]{\keyword{def\;} #1  \keyword{\;=\;}  #2 \keyword{\;in\;} #3}
 
\newcommand*{\gencom}{\com{p}{e}{q}{y}}
\newcommand*{\gensel}{\sel{p}{q}{\ell}}
\newcommand*{\gencond}{\cond{p}{e}{C_1}{C_2}}
\newcommand*{\genrec}{\rec{X}{C_2}{C_1}}

\newcommand*{\apar}{\mathrel{\keyword{\bfseries |}}}
\NewDocumentCommand{\asend}{s m m}
	{{\aspid{#1}{#2}\keyword{!}\langle{#3}\rangle}}
\NewDocumentCommand{\arecv}{s m m}
	{{\aspid{#1}{#2}\keyword{?}#3}}
\NewDocumentCommand{\actor}{s m m}
	{{\aspid{#1}{#2} \actorimpl #3}}
\NewDocumentCommand{\asel}{s m m}{{\aspid{#1}{#2} \mathrel{\color{keyword}\oplus} #3}}
\NewDocumentCommand{\abranch}{s m m}{\aspid{#1}{#2}\mathrel{\keyword{\&}}{#3}}
\newcommand*{\acolon}{\keyword{:}\kern.5ex}

\NewDocumentCommand{\epp}{s m o}{{\left\llbracket{#2}\right\rrbracket\IfNoValueF{#3}{_{\aspid{#1}{#3}}}}}
\def\merge{\sqcup}
\newcommand{\eval}[1]{\downarrow_{#1}}

\NewDocumentCommand{\mqueue}{s m m}{{{\pid m}^{{\aspid{#1}{#2}}}_{{\aspid{#1}{#3}}}}}

\NewDocumentCommand{\rname}{o m}{%
	\ensuremath{\left\lfloor%
	\IfNoValueF{#1}{\text{#1}\middle|}%
	\textsc{#2}%
	\right\rceil}}
\NewDocumentCommand{\rlabel}{m m}{{%
	\customlabel{infrule}{#2}{#1}%
	\hypertarget{#2}{#1}}}
\makeatletter
\NewDocumentCommand{\infrule}{o o m m}{
	\autoflowbox{%
	\begingroup%
	\renewcommand{\arraystretch}{\infrulestretch}%
	\ifdefined\infrule@Nh\else\newlength{\infrule@Nh}\fi
	\ifdefined\infrule@Dh\else\newlength{\infrule@Dh}\fi
	\def\infrule@N{#3}%
	\def\infrule@D{#4}%
	\settoheight{\infrule@Nh}{\mbox{\math\displaystyle\begin{array}{c}\smash{A}\infrule@N\end{array}\endmath}}%
	\settoheight{\infrule@Dh}{\mbox{\math\displaystyle\begin{array}{c}\infrule@D\end{array}\endmath}}%
	\raisebox{\dimexpr \infrule@Nh - \infrule@Dh\relax}{\math\displaystyle%
		\begin{array}{c}
			\infrule@N%
			\\\hline%
			\infrule@D%
		\end{array}%
	\endmath}%
	\endgroup%
	\IfNoValueF{#1}{\IfNoValueTF{#2}{\scalebox{.8}{#1}}%
		{\scalebox{.8}{\rlabel{#1}{#2}}}}%
}}
\makeatother
\newenvironment{infrules}
	{\begin{autoflow}}
	{\end{autoflow}}
\newcommand*{\infrulestretch}{1.3}
\crefname{infrule}{rule}{rules}
\Crefname{infrule}{Rule}{Rules}
\makeatother

\begin{document}

\title{
	\mdseries\scshape\fontsize{18pt}{16pt}\selectfont\MakeLowercase{
	Communications in Choreographies, Revisited
}}

\def\tsa#1#2#3{
	\begin{tabular}{c}
		\href{https://orcid.org/#3}{\Large\scshape\MakeLowercase{#1}}\\[-4pt]
		\small\selectfont\href{mailto:#2}{#2}
	\end{tabular}
	}
\author{
	\tsa{Lu\'\i s Cruz-Filipe}
		{lcf@imada.sdu.dk}
		{0000-0002-7866-7484}
	\tsa{Fabrizio Montesi}
		{fmontesi@imada.sdu.dk}
		{0000-0003-4666-901X}
	\tsa{Marco Peressotti}
		{peressotti@imada.sdu.dk}
		{0000-0002-0243-0480}
	\\[-2pt]\small\normalfont
	Department of Mathematics and Computer Science, University of Southern Denmark
}

\date{}

\twocolumn

\maketitle

\begin{abstract}
Choreographic Programming is a paradigm for developing correct-by-construction concurrent 
programs, by writing high-level descriptions of the desired communications and then synthesising 
process implementations automatically.
So far, choreographic programming has been explored in the \emph{monadic} setting: interaction
terms express point-to-point communications of a single value.
However, real-world systems often rely on interactions of \emph{polyadic} nature, where multiple 
values are communicated among two or more parties, like multicast, scatter-gather, and atomic 
exchanges.

We introduce a new model for choreographic programming equipped 
with a primitive for grouped interactions that subsumes all the above scenarios. Intuitively, 
grouped interactions can be thought of as being carried out as one single interaction. In practice, they
are implemented by processes that carry them out in a concurrent fashion. After formalising the 
intuitive semantics of grouped interactions, we prove that choreographic programs and their 
implementations are correct and deadlock-free by construction.
\end{abstract}

\section{Introduction}
\label{sec:intro}

Choreographic Programming~\cite{M13:phd} is an emerging paradigm for programming communications in 
concurrent and distributed systems. The key idea is that programs are \emph{choreographies}, which define 
the communications that we wish to take place from a global viewpoint, using structures inspired by the 
``Alice-and-Bob'' notation for security protocols~\cite{NS78}.
Then, an \emph{EndPoint Projection} (EPP) synthesises a correct-by-construction implementation in process 
models (\eg process calculi), guaranteeing important properties such as progress and operational 
correspondence~\cite{CHY12,CM13}. The applicability of the paradigm has been demonstrated in 
different settings, including service-oriented programming%
% ~\cite{CHY12,CM13,MY13,chor:website},
~\cite{CHY12,chor:website}% double-blind
, adaptable distributed software~\cite{DGGLM16},
cyber-physical systems~\cite{LNN16,LH17},
and software verification~\cite{CLM17}.

Processes in choreographic programs typically interact via point-to-point message passing. This is 
expressed by language terms like $\gencom$, which reads ``process $\pid p$ evaluates expression $e$ 
locally and sends the result to process $\pid q$, which stores the received value in its local variable 
$y$''.
However, there are application scenarios that require more advanced primitives. We mention two 
representative such scenarios.
First, choreographic languages for cyber-physical systems offer primitives for scatter/gather 
communications (broadcast/reduce in their terminology)~\cite{LNN16,LH17}.
Intuitively, this means generalising $\gencom$ to having many receivers (scatter, $\com*{\pid 
p.e}{\pid q_1.x_1\dots\pid q_n.x_n}$) or many senders (gather, $\com*{\pid q_1.e_1\dots\pid q_n.e_n}{\pid 
p.f}$).
Second, choreographic languages for parallel computing and/or asynchronous communications support the 
idea of exchange~\cite{CLM17}. For example, the term $(\com{p}{x}{q}{y},\com{q}{x}{p}{y})$ 
in~\cite{CLM17} denotes the parallel exchange between processes $\pid p$ and $\pid q$ of their respective 
values locally stored in variable $x$.
These scenarios illustrate the need for choreographic languages with more expressive 
primitives that capture multiple communications. However, the extensions proposed so far differ 
in their syntax and semantics, and none of them comes with an EndPoint Projection procedure. Hence, it is 
still unclear how the correctness-by-construction guarantee of choreographic programming can be 
extended to this kind of primitives.

In this paper, we tackle this issue by extending choreographic programming with language constructs for 
grouping \emph{sets} of communications into complex group interactions, called 
\emph{multicoms}.
Our construct is unifying, in the sense that it captures both the scatter/gather and exchange patterns, 
as we exemplify here. (It actually is even more powerful, as we point out later.)
Consider the following code snippet, a simple program that crawls stores searching 
for the best offer for a given item using the scatter/gather pattern:
\begin{snippet}
	\mcom{
		\com{p}{\code{(item, auth($\pid p$,$\pid s$))}}{s}{t}
	}[\pid s \in S]
	\conc\\
	\mcom{
		\com{s}{\code{priceof($t$)}}{p}{x_{\pid s}}
	}[
		\pid s \in S
	]
\end{snippet}
In the first line, the search service $\pid p$ queries each store $\pid s$ in the collection
$S$ (being sets, multicoms lend themselves to set comprehensions). At first 
sight, this is essentially a multicast as in previous works, but observe that messages from the same 
sender are not necessarily the same: as shown by this example $\pid p$ attaches with information to 
authenticate itself to each receiver. (Hence our primitive is more expressive.) In the second line, 
responses are collected (and possibly aggregated) by $\pid p$. Each step of the interaction between $\pid 
p$ and any given store $\pid s$ is causally dependent, whereas interactions with distinct stores are not 
(request-response interactions with different stores can proceed independently). So, for the first time 
in choreographies, our multicom captures both scatter and gather with a single primitive, but it is not 
limited to those patterns.

Consider now a scenario where two search services, say $\pid p_1$ and $\pid p_2$, run the search code 
above independently and then share their respective offers with each other (one can imagine this to be 
part of a purchase protocol where the service with the best offer then proceeds to buying the item). 
This exchange can be succinctly expressed as the following multicom.
\begin{snippet}
	\resetlinenumber[3]%
	\mcom{
		\com{p_1}{\code{myoffer}}{p_2}{x}\\
		\com{p_2}{\code{myoffer}}{p_1}{x}
	}
\end{snippet}

Programming with multicoms is easy, as we illustrated.
Multicom dynamics should also be straightforward: intuitively, to the programmer's eyes, they are 
interactions among arbitrary groups of processes that are carried out as one.
However, this is not quite what happens in reality, since we know that each communication in a multicom 
may proceed independently from the others whenever possible.
To bridge this gap, we formalise both a simple semantics for choreographies with multicom---where 
statements are run sequentially and multicoms atomically---and a more realistic concurrent 
semantics---where all independent communications may be executed in any possible order. Then, we show that
the concurrent semantics is a refinement of the simple one. This allows for results to be transferred 
between the two semantics. In particular, for the first time, it enables us to extend the 
correct-by-construction methodology of choreographic programming to this kind of structures. Namely, we 
define and prove correct an EndPoint Projection from choreographies to a concurrent process 
calculus.
\section{Choreography Model}
\label{sec:chors}

Typically, there are two kinds of interaction primitives in choreographic programs: \emph{(value) 
communications} and \emph{(interface) selections}~\cite{CM16:facs}.
We shall maintain this distinction for groups of interactions, since each kind 
serves different mechanics, and also to avoid unnecessary technicalities. Concretely, we extend 
choreographies with constructs for grouping communications and for grouping selections called 
\emph{multicoms} and \emph{multisels}, respectively. In this section, we formalise their semantics 
and relevant properties.

\subsection{Syntax}
\label{sec:chors-syntax}

Terms describing choreographic programs (or choreographies, for short) are generated by the 
following grammar:
\begin{align*}
	C \Coloneqq {} &
	\Eta\conc C \mid
	\Phi\conc C \mid
	\gencond \mid
	\genrec \mid
	\\{}&\mid
	X \mid
	\nil
	\\
	\Eta \Coloneqq {} &
		\mcom*{\eta_0\\\dots\\\eta_n}
	\qquad
	\eta \Coloneqq 
%	{} & 
	\com{p}{e}{q}{y}
	\\
	\Phi \Coloneqq {} &
	\msel*{\phi_0\\\dots\\\phi_n}
	\qquad
	\phi \Coloneqq 
%	{} & 
	\sel{p}{q}{\ell}
\end{align*}
A choreography describes the behaviour of a (fixed) set of processes (denoted by $\pid p$,  $\pid 
q$, \etc) running concurrently. Each process has a private memory made of named cells (denoted by 
$x$, $y$, \etc). We assume that each process has a dedicated full-duplex channel to communicate 
with each other process (\eg a TCP/IP channel)---in other words, we assume an underlying 
full-duplex channel for each pair of processes.

Terms $\Eta\conc C$ and $\Phi\conc C$ are (grouped) interactions and read ``the system may execute $\Eta$ (resp.~$\Phi$) and proceed as $C$''. 
An interaction group is either a (finite) set $\Eta$ of value communications ($\eta$-terms like $\gencom$) or a (finite) set $\Phi$ of interface selections ($\phi$-terms like $\gensel$). 
In $\gencom$, $\pid p$ sends its local evaluation of expression $e$ to $\pid q$, which stores the 
received value at $y$. The language of expressions is intentionally not fixed, for 
generality---we just assume that their evaluation always terminates, possibly through a 
timeout.
In $\gensel$, $\pid p$ communicates label $\ell$ (which is a constant) to $\pid q$.
If you like, labels are abstractions of operations, as in service-oriented computing, or 
methods, as in object-oriented computing. So in $\gensel$, $\pid p$ requires $\pid q$ to proceed 
with the behaviour labelled by $\ell$.
In the remainder, we make the standard assumption that choreographies do not contain 
self-interactions (\eg $\sel{p}{p}{\ell}$).

Recall from the introduction that interactions grouped into multicoms and multisels should be 
thought as happening as one or, more precisely, concurrently. This intuition is reflected in the 
fact that multicoms and multisels are sets. As a consequence, interfering interactions cannot be 
grouped. The following is an example of a problematic multicom with interfering interactions:
\begin{snippet*}
\mcom{
	\com{p}{x}{q}{x}\\
	\com{p}{y}{q}{y}\\
	\com{r}{x}{q}{y}\\
	\com{q}{y}{s}{x}
}
\end{snippet*}
In the first two communications, $\pid q$ receives two values from $\pid p$ on the (distinct) 
variables $x$ and $y$. However, the two operations are incompatible: each process can carry out 
actions inside multicoms in any order, but $\pid q$ cannot know which message will arrive first on 
its channel with $\pid p$ unless an order is agreed on (it is not in this case). In the 
implementation of this choreography, it may thus happen that $\pid q$ incorrectly stores in its 
local variable $y$ the value of $x$ at $\pid p$.
The second and third communications are also incompatible: $\pid q$ stores 
the content of the received messages in the same variable $y$, and hence the order in which 
messages are delivered to $\pid q$ cannot be ignored (even though the senders are distinct).
Finally, the third and fourth communications are interfering, too, because the value sent by $\pid 
q$ in the fourth interaction may depend on whether the third interaction takes place before or after it, 
so we cannot interpret the result of the two interactions independently from the order in which 
they are executed.

These observations are formalised by the following syntactic conditions on $\Eta$ terms.
\begin{definition}
	\label{def:multicom}
	A set of communications $\Eta$ is a \emph{(well-formed) multicom} if:
	\begin{enumerate}%[noitemsep,nosep]
	\item if $\mcom{
		\com{p}{e}{q}{y}\\
		\com{r}{e'}{q}{y'}
	} \subseteq \Eta$, then $y \neq y'$ and $\pid p \neq \pid r$;
	\item if $\mcom{
		\com{p}{e}{q}{y}\\
		\com{q}{e'}{r}{y'}
	} \subseteq \Eta$, then $y \notin e'$.
	\end{enumerate}
\end{definition}

Note that even if our model allowed for multiple separate channels between two processes, we would 
still need a requirement like $\pid p \neq \pid r$ in the first condition---\ie we would 
require inequality of channel names, rather than process names.

Similar observations hold for multisels, as illustrated by the following snippet with interfering 
selections:
\begin{snippet*}
\msel{
	\sel{p}{q}{\ell}\\
	\sel{r}{q}{\ell'}\\
	\sel{q}{s}{\ell}
}
\end{snippet*}
Here, process $\pid q$ must concurrently select from an interface at $\pid s$ and await for 
$\pid p$ and $\pid r$ to select an interface each.
\begin{definition}
A set of selections $\Phi$ is a \emph{(well-formed) multisel} provided that:
if $\msel{
		\sel{p}{q}{\ell}\\
		\sel{r}{s}{\ell'}
	} \subseteq \Phi$, then $\pid{q} \notin \{\pid{r},\pid{s}\}$.
\end{definition}
In the remainder, we assume that all multicoms and multisels are well-formed. Also, we often abuse 
notation by writing $\eta\conc C$ and $\phi\conc C$ instead of $\mcom{\eta}\conc C$ and 
$\msel{\phi}\conc C$, respectively.

The remaining choreographic primitives are standard.
In a conditional $\gencond$, the guard $e$ is evaluated in the context of $\pid p$ and then the choreography proceeds executing either branch accordingly---expressions are implicitly assumed to support Boolean values or some equivalent mechanism.
Definitions and invocations of recursive procedures are standard.
The term $\nil$ is the terminated choreography.
As common practice, we assume that all procedure invocations refer to defined procedures, and omit $\nil$ whenever clear from the surrounding terms.

\subsection{Sequential semantics}
\label{sec:chors-big}

\begin{figure*}
	\begin{infrules}
		\infrule[\rname[C]{MCom}][rule:c-mcom]
			{
                                \Eta \neq \emptyset
                                \qquad
				\Eta = \mcom{\com*{\pid p_i.e_i}{\pid q_i.y_i}}[i \in I]
				\qquad
				e_i \eval{\sigma(\pid p_i)} v_i
			}
			{\Eta\conc C, \sigma \to C,\sigma[\pid q_i.y_i \mapsto u_i]}
		\infrule[\rname[C]{MSel}][rule:c-msel]
			{\Phi \neq \emptyset}
			{\Phi\conc C, \sigma \to C,\sigma}
			
		\infrule[\rname[C]{If}][rule:c-if]
			{i = 1 \text{ if } e \eval{\sigma(\pid p)} \literal{true}\text{, } i = 2\text{ otherwise}}
			{\gencond, \sigma \to C_i, \sigma}
	
		\infrule[\rname[C]{Str}][rule:c-struct]
			{C_1 \precongr C_2
				\qquad
				C_2, \sigma \to   C'_2, \sigma'
				\qquad
				C'_2  \precongr C'_1}
			{C_1, \sigma \to  C'_1, \sigma'}

		\infrule[\rname[C]{Ctx}][rule:c-ctx]
			{C_1, \sigma \to C'_1, \sigma'}
			{\genrec, \sigma \to \rec{X}{C_2}{C'_1}, \sigma'}
			
		\infrule[\rname[C]{Unfold}][rule:c-unfold]
			{}
			{\rec{X}{C_2}{C_1[X]} \precongr \rec{X}{C_2}{C_1[C_2]}}
			
		\infrule[\rname[C]{DefNil}][rule:c-defnil]
			{} 
			{\rec{X}{C}{\nil} \precongr \nil}
			
		\infrule[\rname[C]{MEmpty}][rule:c-mempty]
			{} 
			{\mcom{}\conc C \precongr C}						

%		\infrule[\rname[C]{MCom-Com}][rule:c-mcom-com]
%			{} 
%			{\eta \congr \mcom*{\eta}}
%
%		\infrule[\rname[C]{MSel-Sel}][rule:c-msel-sel]
%			{} 
%			{\phi \congr \msel*{\phi}}
	\end{infrules}
%	\vspace{-1ex}
	\caption{Sequential semantics of choreographic programs.}
	\label{fig:chors-big}
%	\vspace{-1ex}
\end{figure*}

We now give a semantics to choreographies, which formalises their intuitive dynamics.
The semantics is the smallest relation $\to$ between pairs $(C,\sigma)$ where
\begin{itemize}%[noitemsep,nosep]
	\item $C$ is a choreography term as defined in \cref{sec:chors-syntax};
	\item $\sigma$ is a function describing the memory of processes (\ie taking processes and 
their variable names to values);
\end{itemize}
that is closed under the rules in \cref{fig:chors-big}. 
Before we discuss each rule observe that:
\begin{itemize}%[noitemsep,nosep]
\item reduction rules consume the outermost interaction (note that the outermost term may be a recursive definition);
\item interactions are consumed in a single step;
\item multicoms are reduced in a single step as if all their interactions were carried out by the involved processes simultaneously.
\end{itemize}
In this sense, this semantics is called sequential or big-step.

For compactness, the presentation relies on the structural precongruence $\precongr$ via the standard mechanism of \crefv{rule:c-struct}; the relation is defined as the smallest relation on choreographic programs closed under \crefv{rule:c-unfold,rule:c-defnil,rule:c-mempty} (discussed below). Herein, $C \equiv C'$ is a shorthand for $C \precongr C'$ and $C' \precongr C$.

The semantics of interactions is defined by \crefv{rule:c-mcom,rule:c-msel}. The expression of each 
communication $\gencom$ in the multicom $\Eta$ is evaluated in the sender context ($e 
\eval{\sigma(\pid p)} v$) and the resulting value ($v$) is used in the \foreign{reductum} to update 
the receiver memory, independently. Selections have no effect on process memory. 
In both cases, the set of communications is required to be non-empty, in order to avoid reductions that do not correspond to any action.
The cases $\Eta=\emptyset$ or $\Phi=\emptyset$ are instead dealt with by structural precongruence (\crefv{rule:c-mempty}).
The semantics of conditionals is modelled by \crefv{rule:c-if}, where the guard is evaluated in the context of the process $\pid p$ performing the choice, and then the program reduces to the corresponding branch.
Recursive definitions are implicitly expanded by structural precongruence, as described by \crefv{rule:c-unfold}: here, $C_1[X]$ indicates that the term $X$ occurs in $C_1$, and on the term on the righthandside it is replaced by the body of the recursive definition. \Crefv{rule:c-ctx} is standard and necessary to reduce the outermost interaction. \Crefv{rule:c-defnil} collects recursive definitions from terminated programs, \ie any $C$ s.t.~$C \precongr \nil$.

\begin{remark} %[Label Selection]
\label{rem:selections}\looseness=-1
Label selections do not alter the state of any process, since they simply model the choice of a 
possible behaviour offered by the receiver. We will use this information to synthesise appropriate 
interfaces for our processes in \cref{sec:epp}. This is a standard method in choreographic 
programming, but we mention it here for the unfamiliar reader.

For example, consider the following choreography.
	\begin{snippet*}
		\cond{p}{e}
		{\\\quad\sel{p}{q}{\textsc{l}}\conc \com{p}{x}{q}{y}\\}
		{\\\quad\sel{p}{q}{\textsc{r}}\conc \com{q}{y}{p}{x}}
	\end{snippet*}
	Here, $\pid p$ makes a local choice and depending on the result selects the appropriate 
behaviour at $\pid q$. In the first case, represented by label $\textsc{l}$, $\pid q$ is expected 
to receive a value from $\pid p$. In the second case, represented by label $\textsc{r}$, $\pid q$ 
is expected to send a value to $\pid p$.
Without label selections, $\pid q$ would not know how to act, since only $\pid p$ 
would know which branch has been selected in the choreographic conditional.
In \cref{sec:epp}, we detail how to synthesise appropriate interfaces for processes such as $\pid 
q$ in this example.
\end{remark}

Choreographic programs never get stuck: they are either successfully terminated or able to reduce.
	
\begin{theorem}
	\label{thm:chors-big-deadlock-freedom}
	For $C$ a choreography, either 
	\begin{enumerate}
		\item $C \precongr \nil$; or,
		\item for every $\sigma$ there are $C'$ and $\sigma'$ such that $C,\sigma \to C',\sigma'$. 
	\end{enumerate}
\end{theorem}

The sequential semantics of choreographic programs enjoys local confluence, which intuitively states that, whenever a computation can proceed in more than one way, it is always possible to reach a common configuration.
Formally:
\begin{theorem}
	\label{thm:chors-big-confluence}
	For every span of computations $C_0,\sigma_0 \to C_1,\sigma_1$ and $C_0,\sigma_0 \to C_2,\sigma_2$ there are $C_3$ and $\sigma_3$ such that $C_1,\sigma_1 \to^\ast C_3,\sigma_3$ and $C_2,\sigma_2 \to^\ast C_3,\sigma_3$, where $\to^\ast$ is the transitive and reflexive closure of $\to$. 
\end{theorem}

\subsection{Concurrent semantics}
\label{sec:chors-small}

\begin{figure*}[t]
	\begin{infrules}
		\infrule[\rname[C]{Com}][rule:c-com]
			{e \eval{\sigma(\pid p_i)} v}
			{\gencom\conc C,\sigma	\to	C, \sigma[\pid q.y \mapsto v]}
		
		\infrule[\rname[C]{Sel}][rule:c-sel]
			{}
			{\gensel\conc C, \sigma \to C, \sigma}
	
		\autoflowbreak
	
		\infrule[\rname[C]{MCom-MCom}][rule:c-mcom-mcom]
			{\Eta_0 = \Eta_1 \uplus \Eta_2}
			{\Eta_0 \congr \Eta_1 \conc \Eta_2}

		\infrule[\rname[C]{MSel-MSel}][rule:c-msel-msel]
			{\Phi_0 = \Phi_1 \uplus \Phi_2}
			{\Phi_0 \congr \Phi_1 \conc \Phi_2}

		\infrule[\rname[C]{MCom-MSel}][rule:c-mcom-msel]
			{
				\mathrm{tn}(\Phi) \cap \mathrm{pn}(\Eta) = \emptyset
			}
			{\Eta\conc\Phi \congr \Phi \conc \Eta}

		\infrule[\rname[C]{MCom-If}][rule:c-mcom-if]
			{
                                x_i \not\in e\mbox{ for all }\com*{\pid q_i.e_i}{\pid p.x_i} \in \Eta
			}
			{
				\cond{\pid p}{e}{(\Eta\conc C_1)}{(\Eta\conc C_2)}
				\congr
				\Eta\conc \cond{\pid p}{e}{C_1}{C_2}
			}

		\infrule[\rname[C]{MSel-If}][rule:c-msel-if]
			{\sel{q}{p}{\ell} \not\in \Phi}
			{
				\cond{\pid p}{e}{(\Phi\conc C_1)}{(\Phi\conc C_2)}
				\congr
				\Phi\conc \cond{\pid p}{e}{C_1}{C_2}
			}

		\infrule[\rname[C]{MCom-Rec}][rule:c-mcom-rec]
			{}
			{\rec{X}{C_2}{(\Eta\conc C_1)} \congr \Eta\conc\genrec}
		\infrule[\rname[C]{MSel-Rec}][rule:c-msel-rec]
			{}
			{\rec{X}{C_2}{(\Phi\conc C_1)} \congr \Phi\conc\genrec}

		\infrule[\rname[C]{If-If}][rule:c-if-if]
			{}
			{
				\begingroup\def\arraystretch{1.1}
				\begin{array}{r}
					\cond{p}{e_1}
						{\cond{q}{e_2}{C_{1}^{1}}{C_{1}^{2}}\\\!}
						{\cond{q}{e_2}{C_{2}^{1}}{C_{2}^{2}}}
				\end{array}
				\endgroup
				\congr
				\begingroup\def\arraystretch{1.1}
				\begin{array}{r}
					\cond{q}{e_2}
						{\cond{p}{e_1}{C_{1}^{1}}{C_{2}^{1}}\\\!}
						{\cond{p}{e_1}{C_{1}^{2}}{C_{2}^{2}}}
				\end{array}
				\endgroup
			}
	\end{infrules}
	\caption{Concurrent semantics of choreographic programs---new rules.}
	\label{fig:chors-small}
\end{figure*}

This section refines the sequential semantics of choreographic programs, redefining the semantics of grouped interactions to allow primitive interactions to proceed independently whenever possible.
When necessary, we distinguish reductions ($\to$) and structural precongruence ($\preceq$) defining the sequential semantics (\cref{sec:chors-big}) and concurrent semantics (\cref{sec:chors-small}) adding subscripts $s$ and $c$, respectively.

The semantics is defined by the rules in \cref{fig:chors-small} together with all rules in \cref{fig:chors-big} except for \crefv{rule:c-mcom,rule:c-msel}.

\Crefv{rule:c-com,rule:c-sel} describe the semantics of primitive interactions between two processes as a specialisation of \crefv{rule:c-mcom,rule:c-msel} to $\mcom{\eta}\conc C$ and $\mcom{\phi}\conc C$, respectively. 
\Crefv{rule:c-mcom-mcom,rule:c-msel-msel} state that groups of interactions can be merged and split 
at runtime---as long as they are well-formed. Merging may not always be possible, since merging 
interactions from distinct groups may violate well-formedness. (In other words, not all interactions 
can be rescheduled and performed concurrently due to causal dependencies.) In the opposite 
direction, it is always possible to split groups, and hence freely schedule their interactions.

In \crefv{rule:c-mcom-msel}, $\mathrm{tn}(\Phi)$ and $\mathrm{pn}(\Eta)$ are the sets of process 
names that occur as selection targets $\{\pid q \mid \gensel \in \Phi\}$ and that occur in a 
communication $\{\pid p, \pid q \mid \gencom \in \Eta\}$, respectively. The rule states that value 
communications and interface selection can be freely scheduled as long as selection targets are not 
involved in any other communication.
\Crefv{rule:c-mcom-if,rule:c-msel-if} state that conditionals and interactions can be swapped as long as the process evaluating the guard is neither the target of an interface selection nor receives a value in a variable that is used by the guard. The remaining rules are straightforward. 

\begin{example}
	Consider the following program:
	\begin{snippet*}
		\mcom{
			\com{p}{e_0}{s_0}{y_0}\\
			\com{p}{e_1}{s_1}{y_1}
		}
		\conc
		\mcom{
			\com{s_0}{e'_0}{p}{x_0}\\
			\com{s_1}{e'_1}{p}{x_1}
		}
	\end{snippet*}\looseness=-1
	Assuming that $e'_i$ depends on $y_i$, this program is a minimal example of the same scatter-gather pattern described in \cref{sec:intro}.
	By immediate applications of \crefv{rule:c-mcom-mcom}, the program is structurally equivalent to \eg:
	\begin{snippet*}
		\mcom{
			\com{p}{e_0}{s_0}{y_0}
		}
		\conc
		\mcom{
			\com{p}{e_1}{s_1}{y_1}\\
			\com{s_0}{e'_0}{p}{x_0}
		}
		\conc
		\mcom{
			\com{s_1}{e'_1}{p}{x_1}
		}
	\end{snippet*}%
	or:
	\begin{snippet*}
		\mcom{\com{p}{e_0}{s_0}{y_0}}
		\conc
		\mcom{\com{s_0}{e'_0}{p}{x_0}}
		\conc
		\mcom{\com{p}{e_1}{s_1}{y_1}}
		\conc\\
		\mcom{\com{s_1}{e'_1}{p}{x_1}}
	\end{snippet*}
	In fact, all these programs yield the very same set of executions: every possible linearisation of the partial order:
	\[\begin{tikzpicture}
		\node[] (n0) {\(\com{p}{e_0}{s_0}{y_0}\)};
		\node[above=.4cm of n0] (n1) {\(\com{s_0}{e'_0}{p}{x_0}\)};
		\node[right=.7cm of n0] (n2) {\(\com{p}{e_1}{s_1}{y_1}\)};
		\node[above=.4cm of n2] (n3) {\(\com{s_1}{e'_1}{p}{x_1}\)};
		
		\draw[thick] (n0) to (n1);
		\draw[thick] (n2) to (n3);
	\end{tikzpicture}
	\tag*{$\triangleleft$}\]
\end{example}

Observe that every (well-formed) multicom is equivalent to any sequence of its interactions:
\[
	\mcom*{\eta_1\\\dots\\\eta_n} \congr \eta_1\conc\dots\conc\eta_n
	\text{.}
\]
Therefore, the semantics of multicoms as per \crefv{rule:c-mcom,rule:c-com} are classified as
big- and small-step---likewise for $\to_{s}$ and $\to_{c}$.
The two are related since the former subsumes the latter (once multicoms are ``sequentialised'') 
and, conversely, the latter subsumes the former once reductions described by \crefv{rule:c-com} are 
transitively aggregated.
The same holds for multisels.
We can generalise these observations into a formal relation between our two semantics.

\begin{lemma}
	\label{thm:chors-big-small-opcorr}
	For any choreography $C_0$ and state $\sigma_0$: 
	\begin{enumerate}%[noitemsep,nosep]
	\item if $C_0,\sigma_0 \to_{s} C_1,\sigma_1$, then $C_0,\sigma_0 \to_{c}^+ C_1,\sigma_1$;
	\item if $C_0,\sigma_0 \to_{c} C_1,\sigma_1$, then there are $C_2$ and $\sigma_2$ s.t.~$C_1,\sigma_1 \to_{c}^\ast C_2,\sigma_2$ and $C_0,\sigma_0 \to_{s}^+ C_2,\sigma_2$;
	\end{enumerate}
	where $(-)^{+}$ is the transitive closure operator.
\end{lemma}

The notion of operational correspondence used in \cref{thm:chors-big-small-opcorr}
is slightly stronger than that studied for process calculi \cite{G10}. The latter is commonly used 
to compare reduction semantics and organise them as refinements and abstractions, but does not 
preserve and reflect progress.
Instead, \cref{thm:chors-big-small-opcorr} immediately allows us to infer progress for $\to_c$.

\begin{theorem}
	\label{thm:chors-small-deadlock-freedom}
	For $C$ a choreography, either 
	\begin{enumerate}
		\item $C \precongr_c \nil$; or,
		\item for every $\sigma$ there are $C'$ and $\sigma'$ such that $C,\sigma \to_c C',\sigma'$. 
	\end{enumerate}
\end{theorem}

Another consequence of \cref{thm:chors-big-small-opcorr} is that the concurrent semantics inherits confluence from the sequential one.

\begin{theorem}
	\label{thm:chors-small-confluence}
	For every span of computations $C_0,\sigma_0 \to_c C_1,\sigma_1$ and $C_0,\sigma_0 \to_c C_2,\sigma_2$ there are $C_3$ and $\sigma_3$ such that $C_1,\sigma_1 \to_c^\ast C_3,\sigma_3$ and $C_2,\sigma_2 \to_c^\ast C_3,\sigma_3$. 
\end{theorem}

\section{Process model}
\label{sec:procs}

In this section, we show that our choreography model supports the correctness-by-construction approach
of choreographic programming. We first introduce our process calculus for modelling concurrent processes, 
and then define an EndPoint Projection (EPP) that synthesises correct process code from choreographies.

\subsection{Syntax}
\label{sec:procs-syntax}

Terms describing process networks are generated by the grammar below.
\begin{align*}
	N
	\Coloneqq {} & 
	\actor{p}{B} \mid \nil \mid N_1 \apar N_2
	\\
	B 
	\Coloneqq {} & 
	\mcom*{\theta_1\\\dots\\\theta_n} \conc B\mid
	\msel{\asel*{\pid q_i}{\ell_i}}_{i \in I}\conc B \mid
	\abranch{\pid p}{\left\{ \ell_i\acolon B_i\right\}_{i\in I}}\mid
	\\ \mid{} &
	\cond*{e}{B_1}{B_2} \mid	
	\rec{X}{B_2}{B_1} \mid
	X \mid
	\nil 
	\\
	\theta
	\Coloneqq {} & 
	\asend{q}{e} \mid
	\arecv{q}{x}
\end{align*}
Networks, ranged over by $N$, are either the inactive network $\nil$, processes $\actor{\pid p}{B}$, where 
$\pid p$ is the name of the process and $B$ its behaviour, or parallel compositions.
A term $\mcom*{\theta_1\\\dots\\\theta_n} \conc B$ represents a behaviour where multiple sends and
receives ($\theta$-terms) are executed concurrently (and thus can be scheduled freely), before proceeding 
with the continuation $B$. In particular, $\asend{\pid q}{e}$ describes a send operation where the 
process evaluates (locally) the expression $e$ and sends the outcome to $\pid q$. Symmetrically, a term 
$\arecv{\pid p}{y}$ represents a receive operation, where the executing process receives a value from $\pid p$ and 
stores it in $y$.
A term $\msel{\asel*{\pid q_i}{\ell_i}}_{i \in I}\conc B$ concurrently sends many label selections 
(each $\ell_i$ is sent to the respective process $\pid q_i$) before proceeding with 
$B$. The dual operation is the branching term $\abranch{\pid p}{\left\{ \ell_i\acolon 
B_i\right\}_{i\in I}}$, where we await to receive from $\pid p$ the selection of one of the labels 
$\ell_i$ and then perform the corresponding behaviour $B_i$.
In a conditional $\cond*{e}{B_1}{B_2}$, the process evaluates the guard $e$ locally and chooses between the continuations $B_1$ and $B_2$ accordingly.
The remaining terms are standard. % (recursive definitions and termination).
We implicitly allow for exchange in the subterms $\mcom*{\theta_1\\\dots\\\theta_n}$,
$\msel{\asel*{\pid q_i}{\ell_i}}_{i \in I}$, and $\left\{ \ell_i\acolon B_i\right\}_{i\in I}$---order does not matter.

\begin{figure*}[!t]
	\begin{infrules}
		\infrule[\rname[P]{Com}][rule:p-com]{
				e \eval{\sigma(\pid p)} v
				\qquad
				\Theta = \mcom{\asend{q}{e}\\\theta_1\\\dots\\\theta_n}
				\qquad
				\Theta' = \mcom{\arecv{p}{x}\\\theta'_1\\\dots\\\theta'_m}
			}{
				\actor{p}{\Theta\conc B} \apar \actor{q}{\Theta'\conc B'},\sigma 
				\to 
				\actor{p}{\mcom{\theta_1\\\dots\\\theta_n}\conc B} \apar \actor{q}{\mcom{\theta'_1\\\dots\\\theta'_m}\conc B'},\sigma[\pid q.y \mapsto v]}
		\infrule[\rname[P]{Sel}][rule:p-sel]{
				g \in I \cap J
			}{
				\actor{p}{\msel{\asel*{\pid q_i}{\ell_i}}_{i \in I}\conc B} \apar 
\actor*{\pid q_g}{\abranch{\pid p}{\{ \ell_j\acolon B_j\}_{j\in J}}},\sigma 
				\to 
				\actor{p}{\msel{\asel*{\pid q_i}{\ell_i}}_{i \in I\setminus \{g\}}\conc B}
				\apar \actor*{\pid q_g}{B_{g}},\sigma
			}
		\infrule[\rname[P]{If}][rule:p-if]
			{i = 1 \mbox{ if } e \eval{\sigma(\pid p)} \literal{true}, i = 2 
\mbox{ otherwise}}
			{\actor{p}{\cond*{e}{B_1}{B_2}},\sigma \to B_i,\sigma}
		\infrule[\rname[P]{Ctx}][rule:p-ctx]
			{\actor{p}{B_1} \apar N,\sigma \to \actor{p}{B_1'} \apar N',\sigma'}
			{\actor{p}{\rec{X}{B_2}{B_1} \apar N,\sigma}
			\to
			\actor{p}{\rec{X}{B_2}{B_1'}} \apar N',\sigma'}
		\infrule[\rname[P]{Par}][rule:p-par]
			{N,\sigma \to  N',\sigma'}
			{N \apar M,\sigma \to N' \apar M,\sigma'}		
		\infrule[\rname[P]{Str}][rule:p-struct]
			{N \precongr M \qquad M \to M' \qquad M' \precongr N'}
			{N \to N'}
		\infrule[\rname[P]{ProcNil}][rule:p-procnil]
			{}
			{\actor{p}{\nil} \precongr \nil}
		\infrule[\rname[P]{ParNil}][rule:p-parnil]
			{}
			{\nil \apar N \precongr N}
		\infrule[\rname[P]{DefNil}][rule:p-defnil]
			{}
			{\rec{X}{B}{\nil} \precongr \nil}
		\infrule[\rname[P]{Unfold}][rule:p-unfold]
			{}
			{\rec{X}{B_2}{B_1[X]} \precongr \rec{X}{B_2}{B_1[B_2]}}
		\infrule[\rname[P]{MEmpty}][rule:p-mempty]
			{}
			{\mcom{}\conc B \precongr B}
	\end{infrules}
	\caption{Process semantics.}
	\label{fig:procs-semantics}
\end{figure*}

\subsection{Semantics}
\label{sec:procs-semantics}

The semantics for process networks is given in \cref{fig:procs-semantics}.
The key difference with respect to the semantics for choreographic programs is that 
execution is now distributed: processes progress concurrently and synchronise only when they interact.
Network semantics is presented in 
terms of a relation between pairs of networks and memory configurations $N,\sigma \to N',\sigma'$.

\looseness=-1
Communication semantics is defined by \crefv{rule:p-com}, which synchronises an output action of a 
process ($\pid p$ in the rule) with an input action at the intended receiver ($\pid q$ in the rule).
Specifically, if there are a ``send-to-$\pid q$'' term ($\asend{q}{e}$) in $\pid p$'s group of 
currently enabled actions and a ``receive-from-$\pid p$'' term ($\arecv{p}{y}$) in $\pid q$'s group of 
currently enabled actions, then $\pid p$ can send the evaluation of $e$ to $\pid q$, and the latter 
updates its local state accordingly ($\sigma[\pid q.y \mapsto v]$).
Similarly, \crefv{rule:p-sel} synchronises a process that wishes to select a branch with the process 
that offers it.
Semantics of conditionals is defined by \crefv{rule:p-if} and is entirely local: the process $\pid p$ performing the choice evaluates the guard $e$ and executes either branch accordingly.
The remaining rules are standard (\cf\ \cite{CM16:facs}).

\subsection{EndPoint Projection}
\label{sec:epp}

\begin{figure*}[t]
	\begingroup
	\newcommand{\afdbox}[1]{\autoflowbox{\math\displaystyle#1\endmath}}
	\makeatletter
		\renewcommand*\autoflow@sep{\qquad}
		\autoflow@middleskip=\smallskipamount
	\makeatother
	\begin{autoflow}
		\afdbox{
			\epp{\nil}[r] \defeq \nil
		}
		\afdbox{
			\epp{\Eta\conc C}[r] \defeq 
			\mcom{\asend{q}{e}\\\arecv{p}{y}}[\com{r}{e}{q}{y}\in \Eta\\\com{p}{e}{r}{y} \in \Eta]\conc\epp{C}[r]
		}			
		\afdbox{
			\epp{\Phi\conc C}[r] \defeq 
			\begin{cases}
				\abranch{\pid p}{\left\{ \ell\acolon \epp{C}[r]\right\}} & \text{if } \sel{p}{r}{\ell} \in \Phi \\
				\msel{\asel{q}{\ell}}[\sel{r}{q}{\ell}\in \Phi]\conc\epp{C}[r] & \text{otherwise}
			\end{cases}
		}
		\afdbox{
			\epp{\gencond}[r] \defeq 
			\begin{cases}
				\cond*{e}{\epp{C_1}[r]}{\epp{C_2}[r]} & \text{if } \pid p = \pid r \\
				\epp{C_1}[r] \merge \epp{C_2}[r] & \text{otherwise}
			\end{cases}
		}
		\afdbox{
			\epp{\rec{X^{\vv{\pid p}}}{C_2}{C_1}}[r] \defeq
			\begin{cases}
				\rec{X}{\epp{C_2}[r]}{\epp{C_1}[r]} & \text{if } \pid r \in \vv{\pid p}\\
				\epp{C_1}[r] & \text{otherwise}
			\end{cases}
		}
		\afdbox{
			\epp{X^{\vv{\pid p}}}[r] \defeq 
			\begin{cases}
				X & \text{if } \pid r \in \vv{\pid p}\\
				\nil & \text{otherwise}
			\end{cases}
		}
	\end{autoflow}
	\endgroup
	\caption{Behaviour projection.}
	\label{fig:epp}
\end{figure*}

Given a choreographic program $C$, we can translate the behaviour of each process $\pid p$ defined in $C$ 
into our process model. We write $\epp{C}[p]$ for this translation, which is defined by structural 
recursion by the rules in \cref{fig:epp}.
All rules follow the intuition of projecting, for each choreography
term, the local action performed by the given process.

Building on $\epp{C}[p]$, we define the EndPoint Projection of a choreography (EPP) as the parallel composition of the behaviours obtained projecting each process separately.
\begin{definition}
	The \emph{EndPoint Projection} (EPP) $\epp{C}$ of a choreography $C$ is the parallel composition:
	\[
		\epp{C} \defeq \prod_{\pid p \in \mathrm{pn}(C)} \actor{p}{\epp{C}[p]}
		\text{.}
	\]
\end{definition}

The key new rules for EPP introduced in this work are the ones for projecting multicoms and multisels.

A multicom term is projected to a group of concurrent send and receive operations, depending on
the role interpreted by the given process in each interaction.
We illustrate this construction in the following example, where we display the choreography that we are projecting on the left and its EPP on the right.
\par\noindent%
\begin{minipage}{.49\columnwidth}
\begin{snippetbox}[height=16mm]\vspace{-1mm}$
	\begin{aligned}%	
	&\mcom*{
		\com{p}{x}{q}{y}\\
		\com{q}{x}{p}{y}
	}\conc\\
	&\mcom*{\com{r}{z}{p}{x}}
	\end{aligned}$
\end{snippetbox}%
\end{minipage}\hfill%
\begin{minipage}{.49\columnwidth}
\begin{snippetbox}[height=16mm]\vspace{-1mm}$
	\begin{aligned}%
	\pid{p}\actorimpl{} & \mcom*{\asend{q}{x}\\\arecv{q}{y}}\conc\mcom*{\arecv{r}{x}}
	\\
	\pid{q}\actorimpl{} & \mcom*{\asend{p}{x}\\\arecv{p}{y}}
	\\
	\pid{r}\actorimpl{} & \mcom*{\asend{p}{z}}
	\end{aligned}$
\end{snippetbox}%
\end{minipage}\par\noindent

\looseness=-1
Multiple selections are handled likewise: a multisel is projected either to a group of selections or to a 
branch, depending on the role of the given process (recall that if a multisel is well-formed, then 
processes cannot occur as selection objects and subjects at the same time).

All remaining rules for EPP are (up to minor differences) standard~\cite{CM16:facs}.
The rules for projecting recursive definitions and calls assume that procedure names have been annotated with the process names appearing inside the body of the procedure, in order to avoid projecting unnecessary procedure code (\cf \cite{CHY12}).

\looseness=-1
The rule for projecting a conditional is more involved.
The (partial) merging operator $\merge$ from \cite{CHY12} is used to merge the 
behaviour of a process that does not know which branch has been chosen yet: $B \merge B'$ is 
isomorphic to $B$ and $B'$ up to branching, where the branches of $B$ or $B'$ with distinct labels 
are also included.
As an example, consider the following choreography
and the projection of its processes.
\par\noindent%
\begin{minipage}{.49\columnwidth}%
\begin{snippetbox}[height=22mm]\vspace{-1mm}$%
	\aligned%
	\cond{p}{e}{%
		{}&	\sel{p}{q}{\textsc l}\conc
		\\ 
		{}&	\com{p}{x}{q}{x}
		\\
	}{
		{}&	\sel{p}{q}{\textsc r}\conc
		\\
		{}& \com{q}{y}{p}{y}
	}
	\endaligned$
\end{snippetbox}
\end{minipage}\hfill%
\begin{minipage}{.49\columnwidth}
\begin{snippetbox}[height=22mm]\vspace{-1mm}$
	\aligned%
	\pid{p}\actorimpl{} & 
		\aligned[t]%
		\cond*{e}{
			{}&\asel{q}{\textsc l}\conc \asend{q}{x}
			\\
		}{
			{}&\asel{q}{\textsc r}\conc \arecv{q}{y}
		}
		\endaligned
	\\
	\pid{q}\actorimpl{} & 
		\abranch{p}{\left\{
			{\textsc l}\acolon \arecv{p}{x}, 
			{\textsc r}\acolon \asend{p}{y}
		\right\}}
	\endaligned$
\end{snippetbox}
\end{minipage}\par\noindent\looseness=-1
Here, merging allows the projection of $\pid q$ to account for the different possible behaviours 
based on the label received from $\pid p$.

If the choreography did not include a selection from $\pid p$ to $\pid q$, then $\pid q$ would not know which choice $\pid p$ had made in evaluating its condition (\cf \cref{rem:selections}). 
This aspect is typical of choreography models \cite{CHY12,CM13,CDYP16,HYC16,DGGLM16,QZCY07}.
More specifically, while the originating choreography executes correctly,
its projection needs processes that behave differently in the branches of a conditional to be 
informed through a selection (either directly or indirectly, by receiving a selection from a 
previously notified process).

Observe that merging is partial and thus there are choreographies whose processes cannot be 
projected. These are not corner cases but actual programming errors that may appear even in simple 
programs like the following one:
\begin{snippet*}
	\cond{p}{e}{\com{p}{e'}{q}{x}}{\nil}	
\end{snippet*}
In this case, the behaviour of process $\pid q$ cannot be projected because $\pid q$ does not know whether 
it should wait for a message from $\pid p$ or not. In general, projections are undefined whenever 
choices operated locally are not correctly propagated to all involved processes; explicit 
selections are thus instrumental to catching such errors at projection time, \ie statically.
Since merging is partial, $\epp{C}[p]$ may be undefined, and consequently $\epp{C}$ is also partial.
In the remainder, we say that a choreography $C$ is projectable if $\epp{C}$ is defined.

\begin{example}
	The projection of the choreographic program discussed in \cref{sec:intro} 
	is the parallel composition of:
	\begin{snippetbox}$
		\begin{aligned}%
		\pid{p}\actorimpl{} & 
			\mcom{\asend{s}{\code{(item, auth($\pid p$,$\pid s$))}}}[\pid s \in S]
			\conc
			\\&
			\mcom{\arecv{s}{x_{\pid s}}}[\pid s \in S]
		\\
		\prod_{\pid s\in S}\pid{s}\actorimpl{} & \arecv{p}{t}\conc\asend{p}{\code{priceof($t$)}}
		\end{aligned}$
	\end{snippetbox}
\end{example}

\subsection{Properties}

\newcommand*\pruning\sqsubset

We end our technical discussion by showing that our framework supports 
the hallmark correctness-by-construction property of choreographic programming.
Formally, this is achieved by proving that a choreography and its EPP are in an operational 
correspondence (they mimic each other); as a corollary, we obtain that the EPP of a choreography is 
deadlock-free.

In our setting, proving an operational correspondence result for EPP is more interesting than in 
previous work on choreographic programming, because we have two semantics for choreographies (the 
sequential relation $\to_s$ and the concurrent relation $\to_c$). Ideally, we would like to get an 
operational correspondence result for each choreographic semantics. A naive way of proceeding would 
be to prove the result twice, once for $\to_s$ and once for $\to_c$. But since we know that $\to_s$ 
and $\to_c$ are related (\cref{thm:chors-big-small-opcorr}), we can do better.

First, we prove the following lemma. We again write $\to^{+}$ for one or more applications of $\to$.
\begin{lemma}
	\label{thm:epp}
	If $C$ is projectable, then:
	\begin{enumerate}
		% Completeness
		\item if $C,\sigma \to_{s} C',\sigma'$ then, there is $N$ such that $\epp{C'}\pruning N$ and ${\epp{C},\sigma} \to^{+} {N,\sigma'}$;
		% Soundness
		\item if ${\epp{C},\sigma} \to {N,\sigma'}$ then, there is $C'$ such that $\epp{C'}\pruning N$ and ${\epp{C},\sigma} \to_{c} {\epp{C'},\sigma'}$.
	\end{enumerate}
\end{lemma}

Above, the pruning relation 
$\pruning$ (from \cite{CHY12,CM13}) drops branches introduced by the 
merging operator $\merge$ when they are no longer needed to follow the originating choreography. 
Pruning is completely orthogonal to our development, so we refer to \cite{CHY12} for a detailed 
explanation.

The choice of $\to_s$ and $\to_c$, respectively, for the two directions in \cref{thm:epp} is 
strategic. Namely, for the first direction, considering $\to_s$ is easier because it is a simpler 
semantics, and it is then straightforward to show that the EPP of the choreography can implement, 
for example, a multicom by executing all its projected process actions. Conversely, for the second 
direction, using $\to_c$ is convenient because it allows us to execute exactly the single move 
performed by the projected network (this may require \eg cherry-picking a single interaction in a 
multicom, or using out-of-order execution for the choreography).

By combining \cref{thm:epp} with \cref{thm:chors-big-small-opcorr}, we get operational 
correspondences for both $\to_s$ and $\to_c$.

\begin{theorem}
	If $C$ is projectable, then:
	\begin{enumerate}
		% Completeness
		\item if $C,\sigma \to_{s} C',\sigma'$ then, there is $N$ such that $\epp{C'}\pruning N$ and 
${\epp{C},\sigma} \to^{+} {N,\sigma'}$;
		% Soundness
		\item if ${\epp{C},\sigma} \to {N,\sigma'}$ then, there are $N'$, $C'$, and $\sigma''$ such 
that $\epp{C'}\pruning N'$, ${C,\sigma} \to_{s}^{+} {C',\sigma''}$, and ${N,\sigma'} \to^{\ast} 
{N',\sigma''}$.
	\end{enumerate}
\end{theorem}

\begin{theorem}
	If $C$ is projectable, then:
	\begin{enumerate}
		% Completeness
\item if ${C,\sigma} \to_{c} {C',\sigma'}$ then, there are $N$, $C''$, and $\sigma''$ such that 
$\epp{C''}\pruning N$, ${C',\sigma'} \to_{c}^{\ast} {C'',\sigma''}$, and ${\epp{C},\sigma} \to^{+} 
{N,\sigma''}$;
		% Soundness
		\item if ${\epp{C},\sigma} \to {N,\sigma'}$ then, there is $C'$ such that $\epp{C'}\pruning 
N$ and ${\epp{C},\sigma} \to_{c} {\epp{C'},\sigma'}$.
	\end{enumerate}
\end{theorem}

As a corollary of the operational correspondences that we developed and the progress property of choreographies we get that projected networks are deadlock-free.

\begin{corollary}
	Let $N = \epp{C}$ for some $C$. Either 
	\begin{enumerate}
		\item $N \precongr \nil$ ($N$ has terminated), or 
		\item for any $\sigma$ there exist $N'$ and $\sigma'$ such that $N,\sigma \to N',\sigma'$ ($N$ can always reduce).
	\end{enumerate}
\end{corollary}

\section{Related work and Conclusions}
\label{sec:extensions-and-related}

Scatter/gather primitives for choreographic programs were introduced in~\cite{LNN16}, 
in order to use choreographies for modelling cyber-physical systems.
The primitive of asynchronous exchange, where two processes exchange a value at the same time, 
was introduced in~\cite{CLM17}. Neither of these works discussed how such primitives may be 
supported in the paradigm of Choreographic Programming~\cite{M13:phd}, in order to generate 
correct-by-construction implementations. Furthermore, these primitives are not general, in the 
sense that one construct cannot be used to obtain the same effect as the other.
In this work, we have addressed both issues, by introducing unifying primitives (our 
multicoms/multisels) that can capture both patterns and defining an EndPoint Projection that 
generates correct process terms in a calculus of concurrent processes.

Some previous works on choreographic programming includes a parallel composition operator for 
choreographic terms ($C \apar C'$), for example \cite{CHY12}. Implementing our multicom/multisel 
using parallel composition requires the possibility to join the two terms $C$ and $C'$ after they 
have finished execution, and then to proceed with a continuation. These are not supported in 
\cite{CHY12}. Instead, both a parallel composition operator and a general sequential composition 
operator ($C;C'$) are present in \cite{DGGLM16}, which would in theory allow for encoding our 
grouped interactions. However, the combination of these two operators can cause EPP to generate 
interfering communication actions between the parallel branches, and between the parallel branches 
and the continuation. A correct EPP in \cite{DGGLM16} is then obtained by adding (i) distinct 
auxiliary communication channels for communications and (ii) hidden communications for the 
propagation of information about internal choices at participants. Our approach is more efficient, 
because (i) we simply assume one reusable duplex channel for each pair of processes (used by all 
communications between them) and (ii) our well-formedness condition for multisels combined with 
merging guarantees that all participants agree without the need for hidden communications. 
Furthermore, the model in \cite{DGGLM16} does not consider our well-formedness conditions for 
multicoms and may thus lead to confusing data races. For example, the (equivalent of the) exchange 
$\{ \com{p}{x}{q}{x},\com{q}{x}{p}{x} \}$ is allowed (notice that $x$ is used both for receiving and 
sending), which in a synchronous system would \emph{never} yield the expected exchange, but rather a 
copy of $x$ from $\pid p$ to $\pid q$ or vice versa (since one of the two interactions must be fully 
performed before the other can).

Most works on choreographic programming fall into two categories: those that use a 
sequential semantics (like \cite{LGMZ08,CHY12,CMS17,DGGLM16}), and those that
allow for out-of-order concurrent execution (like \cite{CM13,MY13,CM16:facs,LNN16}). So far, 
adopting the first view meant requiring the programmer to write all concurrent behaviour manually, 
which could be error-prone (\cf the complex verification techniques for detecting some mistakes in 
\cite{CHY12}). And, adopting the second view meant sacrificing the straightforward semantic 
interpretation of choreographies. Our development bridges this gap and offers a 
third view: programmers can use the sequential semantics of choreographies to reason about 
communication behaviour---in a language where concurrency does not need to be manually specified, 
because we simply abstract from it---and then stand on the shoulders of our results for the 
concurrent semantics (by operational correspondence) to know that safety is preserved in concurrent 
implementations.
This result has an important practical implication: it is feasible to build a debugger for 
choreographies that uses the sequential semantics, since all results will be equivalent 
anyway---this would help programmers, since they would have to debug many fewer possible 
executions. But we do not need to give up the efficiency and realism of the concurrent semantics 
for runtime process implementations.

In \cite{CM17:ice}, an operational correspondence result is presented in the 
setting of asynchronous communications for the calculus of core choreographies \cite{CM16:facs}; 
specifically, the authors show that programmers can interpret core choreographies as using 
synchronous communications, and that there is a safe way of obtaining more efficient asynchronous 
implementations without needing manual intervention.
We conjecture that our development may be combined with that in \cite{CM17:ice},
to obtain an asynchronous semantics for grouped interactions. We leave this combination to future 
work.

The congruence rules for swapping independent choreographic interactions were first introduced 
in~\cite{CM13}; we have extended them to deal with groups of interactions (multi\-coms/multi\-sels). 
Our terms for recursion and conditionals in choreographies are standard, 
from~\cite{CHY12,CM16:facs}. Likewise, the terms for recursion, conditionals, and parallel 
composition of networks in our process model are borrowed from~\cite{CM16:facs}.

\paragraph{Acknowledgements}
	This work was partially supported by the Open Data Framework project at the University of Southern Denmark, and by the Independent Research Fund Denmark,
	Natural Sciences, grant DFF-7014-00041.

\bibliography{biblio}

\end{document}